\documentclass{PHYSEAUTH}
\usepackage{graphicx}
\usepackage{amsmath}
\usepackage{amssymb}

\begin{document}

\begin{frontmatter}

\title{
Formulation and Application of Quantum Monte Carlo Method
to Fractional Quantum Hall Systems
}

\author[address1]{Sei~Suzuki\thanksref{thank1}}
and
\author[address2]{Tatsuya~Nakajima}

\address[address1]{Department of Basic Science, University of Tokyo, 
Tokyo, 153-8902, Japan}

\address[address2]{Department of Physics, Tohoku University, 
Sendai, 980-8578, Japan}

\thanks[thank1]{
Corresponding author. 
E-mail: sei@sola.c.u-tokyo.ac.jp}

\begin{abstract}
Quantum Monte Carlo method is applied to fractional
quantum Hall systems. 
The use of the linear programming method enables us to avoid 
the negative-sign problem in the Quantum Monte Carlo calculations. 
The formulation of this method 
and the technique for avoiding the sign problem are described. 
Some numerical results 
on static physical quantities are also reported.
\end{abstract}

\end{frontmatter}

\section{Introduction}
Two-dimensional electron gas (2DEG) in a strong 
magnetic field is known as a system 
which shows various phenomena resulting from electronic interactions. 
In particular, the fractional quantum Hall (FQH) effect 
has attracted much theoretical and experimental attention 
as a correlation-dominated one peculiar to this system.
To investigate such strongly correlated systems theoretically, 
numerical studies play a significant role.
For example, the exact diagonalization and density matrix 
renormalization group (DMRG) methods~\cite{DMRG} have been used 
for the study of FQH systems, 
because the ones are directly accessible to the ground state. 
However, the exact diagonalization can 
study static and dynamical properties only for small systems, 
while it is not so easy to obtain dynamical information 
by 
the DMRG calculations.
Thus 
we apply
the quantum Monte Carlo (QMC) method to this system
in the present work.

The QMC method has been used for several quantum 
many-body systems.
It can investigate both static and dynamical properties
of systems larger than those to which 
the exact diagonalization is applicable.
However, this method is inherently 
involved with the so-called 
negative-sign problem.
Thus we need to overcome (at least, or moderate) 
the negative-sign 
nuisance 
in order to make the best use of the method.
In this study, we 
formulate the QMC method to the FQH system 
that is free of the sign problem 
and report some numerical results.


\section{General formalism of QMC at zero temperature}
In the zero-temperature formalism, 
the expectation value of an observable $\mathcal{O}$ for a 
Hamiltonian $\mathcal{H}$ is given by
\begin{equation}
 \langle\mathcal{O}\rangle = 
\lim_{\beta\to\infty} \langle\psi| e^{-\beta\mathcal{H}/2} 
\mathcal{O} 
e^{-\beta\mathcal{H}/2}|\psi\rangle /
\langle\psi| e^{-\beta\mathcal{H}} |\psi\rangle,
\label{eq:<O>}
\end{equation}
where $|\psi\rangle$ is an arbitrary state not orthogonal to
the ground state. 
Here we assume that the Hamiltonian can be written in
a quadratic form of a one-body operator.
When the imaginary-time evolution operator 
$e^{-\beta\mathcal{H}}$ is decomposed to 
imaginary-time slices $e^{-\Delta\beta\mathcal{H}} \cdots 
e^{-\Delta\beta\mathcal{H}}$, 
we perform the Hubbard-Stratonovich (HS) transformation
for the quadratic form by introducing auxiliary fields 
for each time slice.
Then the Hamiltonian is reduced to 
a linear form of the one-body operator
and the expectation value is expressed 
in a form of an auxiliary-field path integral. 
We evaluate it by means of the Monte Carlo method.
However, statistical weights used in this Monte Carlo calculation 
are not always positive for any auxiliary-field configuration
and then the normalization denominator in Eq.(\ref{eq:<O>})
often becomes almost zero.
Thus it is often impossible to 
calculate $\langle\mathcal{O}\rangle$ with sufficient precision.
This disaster is called the negative-sign problem.


\section{Hamiltonian}
We consider interacting electrons confined on a spherical surface 
with a magnetic monopole located at the center of the
sphere~\cite{Sphere}.  
For simplicity, we neglect the spin degrees of freedom 
and assume that all the electrons occupy the lowest Landau level.
Then single-particle states are specified by the $z$-component, $m$, 
of angular momentum whose amplitude is $s$, where $2s$ is the number of
flux quanta piercing the sphere and $m$ ranges from $-s$ to $s$.
The Hamiltonian can be written in a quadratic form of 
the density operator and its time reversal:
\begin{equation}
 {\mathcal H} = - \frac{1}{2}\sum_{K=0}^{2s}\chi_K\sum_{N=-K}^K
 \rho_{_{K N}}\tilde{\rho}_{_{K N}} + C_0 \,\rho_{_{0 0}} ,
 \label{eq:H}
\end{equation}
\[
 \rho_{_{K N}} \equiv 
\sum_{m_1, m_2}
  \langle 
  K N | m_1 m_2 \rangle (-1)^{s+m_2}\,
 a^{\dagger}_{m_1} a_{-m_2}, ~~~
\]
where $a_m$ is an annihilation operator of electron, 
$\langle K N |m_1 m_2 \rangle$ the Clebsch-Gordan coefficient,
and $C_0$ is a constant.
The time reversal of the density operator is defined by
$\tilde{\rho}_{_{K N}} = (-1)^{K}\,\rho^{\dagger}_{_{K N}}$.
The coupling constant $\chi_K$ is related with the Haldane 
pseudopotential $V_J$ as 
\begin{equation}
 \chi_K = \sum_{J=0}^{2s} T_{K J} V_J, \label{eq:chi}
\end{equation}
\[
 T_{K J} = (-1)^{2s-J+K+1}(2J+1)
 \begin{Bmatrix}
  s & s & J \\
  s & s & K
 \end{Bmatrix},
\]
where the braces denote Wigner's
$6j$ symbol.
Since the transformation matrix $T$ satisfies $T^{-1}=T$,
we also have a relation of 
\begin{equation}
V_J = \sum_{K=0}^{2s}T_{J K}\chi_K.
 \label{eq:V}
\end{equation}

The Hamiltonian in Eq.(\ref{eq:H}) can be written 
in a linear form of the density operator by the HS transformation. 
Denoting the auxiliary field for a mode $(K N)$ by $\sigma_{K N}$, 
the linearized Hamiltonian is given by
\begin{eqnarray*}
 h(\sigma) &=& \sum_{K=1}^{2s}\eta_K\chi_K 
 \left[\sum_{N=1}^K
 (\sigma_{K N}^{\ast}\rho_{K N} + \sigma_{K N} \tilde{\rho}_{K N}) 
 \right. \\
 & & \left. + \frac{1}{2}(\sigma_{K 0}^{\ast}\rho_{K 0} 
 + \sigma_{K 0}\tilde{\rho}_{K 0}) 
 \right]
 + \frac{1}{2}\chi_0\rho_{00}^2 + C_0\rho_{00},
\end{eqnarray*}
where 
$\eta_K = i$ when $\chi_K < 0$ and $\eta_K = 1$ for 
non-negative $\chi_K$.
We note that $\rho_{00}$ is proportional to the number operator 
of electrons. 
Since the number of electrons is a conserved quantity 
in the zero-temperature formalism, 
the term $\chi_{0}\rho_{00}\tilde{\rho}_{00}$ and 
$C_0 \,\rho_{_{0 0}}$ in Eq.(\ref{eq:H}) can be considered as constants.

\section{Sign problem}
In order to discuss the sign problem, we formulate here 
the auxiliary field path integral in a matrix representation. 
We first define the matrix elements of the
linearized Hamiltonian by
\begin{equation}
 -\Delta\beta \,h(\sigma) \,= \sum_{m,n=-s}^s M_{m,n}(\sigma)\,
 \alpha_m^{\dagger}\alpha_n,
 \label{eq:M}
\end{equation}
where 
\[
 \alpha_m =  \left\{\begin{array}{@{\,}ccl@{\,}}
   a_m & (m\ge 0),  \\
   (-1)^{s-m}a_m& (m<0).  \end{array}
 \right. 
\]
Next, we represent the state $|\psi\rangle$ in Eq.(\ref{eq:<O>})
in terms of a matrix $V$ as
\[
 |\psi\rangle = (\sum_{m}\alpha_{m}^{\dagger}V_{_{m 1}})
 (\sum_{m}\alpha_{m}^{\dagger}V_{_{m 2}}) \cdots
 (\sum_{m}\alpha_{m}^{\dagger}V_{_{m N_e}})|0\rangle ,
\]
where $N_e$ is the number of electrons. We remark that 
$M$ is a square matrix of dimension $2s+1$ and $V$ is a
$(2s+1)\times N_e$ rectangular matrix. Then the denominator of
the Eq.(\ref{eq:<O>}) is written as~\cite{QMCBook}
\[
 \langle\psi|e^{-\beta\mathcal{H}}|\psi\rangle =
 \int\mathcal{D}\sigma \,G(\sigma)\,\zeta(\sigma),
\]
\[
 \mathcal{D}\sigma \equiv\prod_i \prod_{K=1}^{2s}\prod_{N=0}^K
 d^2\sigma_{KN}^{\left(i\right)} ,
\]
\begin{eqnarray*}
 G(\sigma) &\equiv& \prod_i 
 \left \{
  \prod_{K=1}^{2s}
 \biggl[
 \frac{\Delta\beta|\chi_{K0}|}{4\pi}
 \prod_{N=1}^K \biggl(
 \frac{\Delta\beta|\chi_{KN}|}{2\pi} \biggl)
  \biggl] \right.\\
 &\times& \left. \exp 
 \biggl[
 -\Delta\beta\sum_{K=1}^{2s} |\chi_K|
  \biggl(\sum_{N=1}^K |\sigma_{KN}^{\left(i\right)}|^2+
  \frac{|\sigma_{K0}^{\left(i\right)}|^2}{2}\biggl)
  \biggl] 
  \right \},
\end{eqnarray*}
\begin{eqnarray}
 \zeta(\sigma) &\equiv& \langle\psi|
 e^{-\Delta\beta h(\sigma^{\left(N_t\right)})}\cdots
 e^{-\Delta\beta h(\sigma^{\left(2\right)})}
 e^{-\Delta\beta h(\sigma^{\left(1\right)})}
 |\psi\rangle \nonumber \\
 &=& \det\left[V^{\dagger}e^{M(\sigma^{(N_t)})}\cdots
	 e^{M(\sigma^{(2)})} e^{M(\sigma^{(1)})}V\right],
 \label{eq:zeta}
\end{eqnarray}
\noindent
where $N_t = \beta/\Delta\beta$ is the number of the
imaginary-time slices.

The sign problem is brought about 
by the fact that $\zeta(\sigma)$ can be negative. 
However 
$\zeta(\sigma)$ is forced to be non-negative under the following 
conditions~\cite{SMMCPRC}: \\
(i) $2s$ is odd and $N_e$ is even, \\
(ii) the coupling constants in the Hamiltonian
satisfy $\chi_K\ge 0$ for $K = 1, 2, \cdots, 2s$.\\
We show this as follows. From the condition (i), 
the indices, $m$ and $n$, of
the matrix $M(\sigma)$ in Eq.(\ref{eq:M}) take the values, 
$\pm\frac{1}{2}, \pm\frac{3}{2}, \cdots, \pm s$. 
The matrix $M(\sigma)$ under the above conditions satisfies 
two relations as 
$M_{-m,-n}(\sigma) = M_{m,n}^{\ast}(\sigma)$ and 
$M_{-m,n}(\sigma) = - M_{m,-n}^{\ast}(\sigma)$ 
for positive $m$ and $n$.
When the matrix $V$ satisfies $V_{-m,N_e/2+i}=V_{m,i}^{\ast}$
and $V_{-m,i}=-V_{m,N_e/2+i}^{\ast}$ for positive $m$ and
$i=1,\cdots,N_e/2$, then the matrix in 
Eq.(\ref{eq:zeta}) can be written as
\[
 V^{\dagger}e^{M(\sigma^{(N_t)})} 
 \cdots e^{M(\sigma^{(1)})}V
 = \left[\begin{array}{@{\,}cc@{\,}}
   A & B  \\
   -B^{\ast} & A^{\ast}  \end{array}\right],
\]
where $A$ and $B$ are square matrices of dimension $N_e/2$.
This type of matrix always has paired eigenvalues that are 
complex-conjugate each other.
In fact, 
denoting a matrix of the above form and its eigenvalue 
by $X$ and $\lambda$, respectively, 
the complex conjugation of the proper equation
yields the one on $\lambda^{\ast}$:
\begin{eqnarray*}
 0 
 &=& \det \left[
   \left[\begin{array}{@{\,}cc@{\,}}
   {\bf 0} & -{\bf 1}  \\
   {\bf 1} & {\bf 0}  \end{array}\right]
   X
   \left[\begin{array}{@{\,}cc@{\,}}
   {\bf 0} & {\bf 1}  \\
   -{\bf 1} & {\bf 0}  \end{array}\right]
    - \lambda^{\ast}\right] 
 = \det \left[
  X - \lambda^{\ast}\right] . \nonumber 
\end{eqnarray*}
That is, $\lambda^{\ast}$ is also an eigenvalue of $X$
and thus the determinant of this type of matrix
is always non-negative, 
which leads to the desired condition that  
$
 \zeta(\sigma) \ge 0
$.

Now we consider which FQH systems satisfy the above conditions.
The condition (i) can be satisfied, for example, 
in case of $2s = 3N_e-3$ for the $\nu=1/3$ Laughlin state~\cite{Sphere}, 
or $2s = 2N_e-3$ for the Pfaffian state~\cite{Pfaffian}.
However, the condition (ii) is not satisfied when the values 
for the Coulomb interaction are used for $V_J$ in Eq.(\ref{eq:chi}).
Thus we need to control the value of $\chi_K$ (that is, that of $V_J$) 
by solving a linear programming problem 
in order to satisfy the condition (ii). 
Namely,
taking into account that only $V_J$ for 
odd $2s-J$
are physical for fermionic systems,
we inquire 
which set of $\chi_K$ minimizes 
\begin{equation}
 F(\chi_0,\chi_1,\cdots,\chi_{2s}) \equiv
\sum_{2s-J:{\rm odd}}\lambda_J(\sum_{K=0}^{2s}T_{J K}\chi_K - V_J )
\label{eq:Linear}
\end{equation}
under the conditions:
$\chi_0 \le 0$, 
$\chi_K \ge 0$ for 
$K=1,2,\cdots,2s$,
and 
$\lambda_J (\sum_{K=0}^{2s}T_{J K}\chi_K - V_J ) \ge 0$
for odd $2s-J$.
We note here that 
$\lambda_J$ controls the 
variance of pseudopotential from that 
for the Coulomb interaction for each $J$.
In the present study we always set $\lambda_J = 1$.
Negative $\chi_0$ is needed to make 
$V_0$ ($=\sum_K T_{0K}\chi_K$) positive
under the fact that $T_{0K}<0$ for all $K$.
Thus obtained $\chi_K$ minimizes the variance from that for 
the Coulomb interaction satisfying the condition (ii).

\section{Numerical results}
\begin{figure}
\begin{center}\leavevmode
 \includegraphics[width=7.5cm,clip]{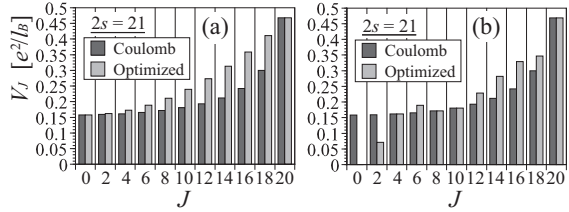}
\end{center}
\caption{Pseudopotentials for the Coulomb interaction 
and negative-sign-free potentials obtained by 
 the linear programming method for $2s=21$.
 The referred potential in the linear programming method 
 is the Coulomb potential in (a), but in (b) 
 long-range components, $V_0$ and $V_2$, of the referred potential
 are set to be zero.
 Only the physical ($2s-J$: odd) components are shown.
}
\label{fig:Potentials}
\end{figure}

Figure~\ref{fig:Potentials} (a)
shows an example of pseudopotentials free of the negative-sign problem.
Although the optimized potentials do not coincide completely 
with those for the Coulomb interaction, 
their monotonical dependence on $J$ is realized naturally.
We show another example of sign-problem free potential in 
Fig.\ref{fig:Potentials}(b). 
To get the short-range components closer
to those for the Coulomb potential, 
we bring long-range components close to zero. 
Namely, we set $V_0 = V_2 = 0$ in Eq.(\ref{eq:Linear}).
As a result, variance of short-range components 
from those of the Coulomb 
potential is squeezed. 
We note that optimized pseudopotentials can be obtained 
with less variance for higher Landau levels or 
in the presence of finite-thickness effects.


\begin{figure}
\begin{center}\leavevmode
 \includegraphics[width=7.5cm,clip]{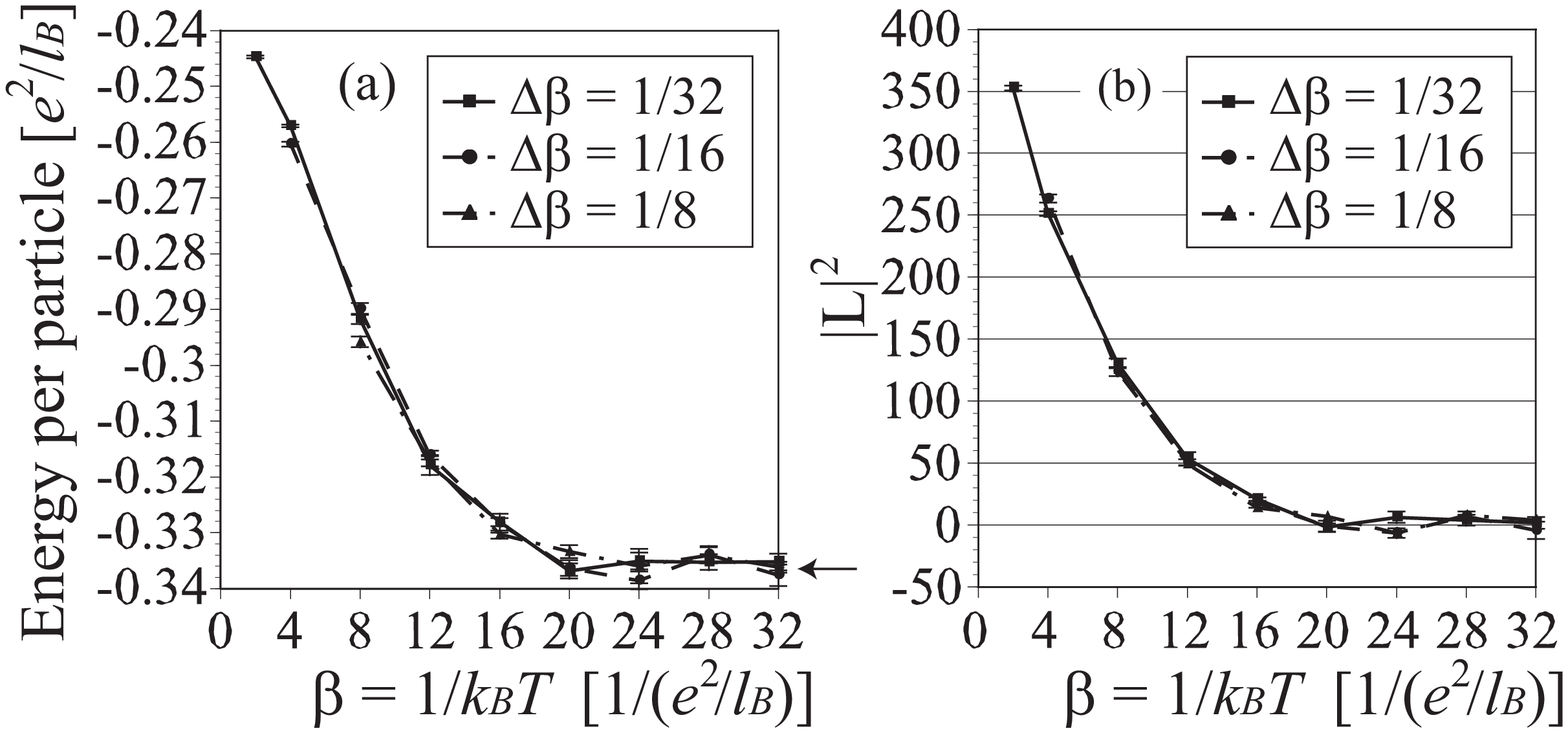}
\end{center}
 \caption{
 Expectation values of (a) energy per particle and 
 (b) total angular momentum by the QMC calculation
 are shown versus inverse temperature
 for $2s=21$ and $N_e=8$.
 Optimized potential in Fig.\ref{fig:Potentials}(b) was used.
 The values used as imaginary-time slice width are 
 $\Delta\beta=1/8$, $1/16$, and $1/32$ in units of $1/(e^2/l_B)$. 
 The ground state energy obtained by the exact diagonalization is shown
 by an arrow on the right side of (a).
 }
 \label{fig:Energy}
\end{figure}

In Figure~\ref{fig:Energy},
we show expectation values of energy and total
angular momentum 
against 
the inverse temperature $\beta$.
Although finite value of $\Delta \beta$ gives rise to 
numerical errors in the Suzuki-Trotter decomposition,
expectation values almost saturate
for $\Delta\beta \le 1/8\,(e^2/l_B)$
($l_B \equiv \sqrt{c /e B}$: the magnetic length).
For comparison, we show the ground state energy obtained by 
the exact diagonalization by an arrow in Fig.\ref{fig:Energy}(a).
It is noted that the total angular momentum of 
the true ground state is zero.

We find that for $\beta > 20/(e^2/l_B)$ 
these two quantities converge to the 
values expected for the ground state.
In fact, 
the value of energy gap is about $0.1 e^2/l_B$
in our exact diagonalization study, and thus 
the $\beta$-value choice of $\beta > 20$ is 
appropriate to investigate the ground state properties.
In case of smaller energy gap,
larger values of $\beta$ are needed for the convergence 
of expectation values. 
Thus incompressible gapful states are more suitable
 than gapless ones for the study by the present method.

%
%
%
%
It is significant to discuss 
whether the ground state for negative-sign-free
interaction is similar to that for the Coulomb interaction.
Unfortunately the overlap between these two states for $\nu = 1/3$
is not so large ($\sim 0.38$)
for the pseudopotentials shown in Fig.\ref{fig:Potentials}(b).
Since the negative-sign-free interaction
with $V_{2s-1}$ quite larger than $V_{2s-3}$ 
is expected to stabilize the Laughlin state 
as the lowest-energy state, 
the small overlap value seems to
come from the reduction in long-range components.
However, taking into account the fact that 
the ground state is gapful and the overlap is finite, 
the ground state for the negative-sign-free interaction 
is considered to capture essential properties of the Laughlin state.

\section{Acknowledgement}
One of the authors (S.S.) acknowledges support by Research Fellowship 
for young scientists of JSPS.
The present work is supported by Grant-in-Aid 
for Scientific Research (Grant No.1406899 and No.14740181) 
by the Ministry of Education, Culture, Sports, Science and 
Technology of Japan.

\end{document}